\documentclass[epsfig,aps,12pt]{revtex4}
\usepackage{amsmath, amssymb}
\def\N{\hbox{${\cal N}$~}}
\begin{document}

\title{A braneworld puzzle about entropy bounds and a maximal temperature  }

\author{Ram Brustein \protect\( ^{(1)}\protect \), David Eichler \protect\( ^{(1)}\protect \),
and Stefano Foffa \protect\( ^{(2)}\protect \)}
\affiliation{(1)
Department of Physics, Ben-Gurion University, Beer-Sheva 84105,
Israel
\\
(2) Department of Physics,  University of Geneva, CH-1211 Geneva,
Switzerland\\
 \texttt{e-mail:  ramyb,eichler@bgumail.bgu.ac.il },
\texttt{Stefano.Foffa@physics.unige.ch}}

\begin{abstract}

Entropy bounds applied to a system of \N species of light quantum
fields in thermal equilibrium at temperature $T$ are saturated in
four dimensions at a maximal temperature
$T_{max}=M_{Planck}/\sqrt{\N}$. We show that the correct setup for
understanding the reason for the saturation is a cosmological
setup, and that a possible explanation is the copious production
of black holes at this maximal temperature. The proposed
explanation implies, if correct, that $\N$ light fields cannot be
in thermal equilibrium at temperatures $T$ above $T_{max}$.
However, we have been unable to identify a concrete mechanism that
is efficient and quick enough to prevent the universe from
exceeding this limiting temperature. The same issues can be
studied in the framework of AdS/CFT by using a brane moving in a
five dimensional AdS-Schwarzschild space to model a radiation
dominated universe. In this case we show that $T_{max}$ is the
temperature at which the brane just reaches the horizon of the
black hole, and that entropy bounds and the generalized second law
of thermodynamics seem to be violated when the brane continues to
fall into the black hole. We find, again, that the known physical
mechanisms, including black hole production, are not efficient
enough to prevent the brane from falling into the black hole. We
propose several possible explanations for the apparent violation
of entropy bounds, but none is a conclusive one.

\end{abstract}
\pacs{PACS numbers: ??.??.??}

\maketitle
\section{Introduction}
\label{intro}

Entropy bounds seem to imply that \N light quantum fields cannot
be in thermal equilibrium at an arbitrarily high temperature. In
four dimensions they are saturated at a temperature equal to
$T_{MAX}= M_P/ \N^{1/2}$ (here $M_P$ is the Planck mass). When
entropy bounds are saturated it is possible, in many cases, to
identify a physical mechanism that enforces them. The prime
candidate for such a mechanism is black hole (BH) production. If
many BH's are produced, the system goes into a kind of phase
transition. In the new phase the previous energy and entropy
estimates are no longer valid. Since BH's are more efficient in
storing entropy, the bounds are not violated.

We seek a physical mechanism that places an upper bound on the
temperature, if such an upper bound indeed exists. Since we wish
to use semi-classical methods and avoid the quantum regime, we
focus on the limit of large \N since then $T_{MAX} \ll M_P$. As we
will show, the correct context for studying this issue is a
cosmological context.

Previously, Bekenstein \cite{beb1} argued that if the entropy of a
visible part of the universe obeys the usual entropy bound from
nearly flat space situations \cite{beb2}, then the temperature is
bounded and therefore certain cosmological singularities are
avoided. More recently, there have been several discussions
following a similar logic.  Veneziano \cite{veb} suggested that
since a BH larger than a cosmological horizon cannot form
\cite{ch}, the entropy of the universe is always bounded. This
suggestion is related, although not always equivalent, to the
application of the holographic principle \cite{holo} in cosmology
\cite{fs,EL,KL,entbound1,entbound2,entbound3,entbound4,entbound5,entbound6,entbound7}.
In \cite{gsl,gslstring,Brustein:2001di} it was argued that the
Hubble parameter $H$ is bounded by entropy considerations, $ H\le
H_{MAX}\equiv\frac{M_P}{\sqrt{\N}}$. In a cosmological context
this is equivalent to $T\le T_{MAX}$.

The AdS/CFT correspondence \cite{Maldacena:1997re,adsrev} offers
an alternative route and a new perspective for the study of a
system of a large number \N of light fields in thermal equilibrium
in a cosmological setup by studying brane propagation in an
AdS-Schwarzschild background \cite{Kraus:1999it,
Kehagias:1999vr,Ida:1999ui,Wang1,Wang2}. Branes moving in
AdS-Schwarzschild space are expected to be dual to finite
temperature CFT's in a cosmological background
\cite{Witten:1998zw, Gubser:1999vj}. However, the status of the
conjecture is somewhat weaker than the one relating to an AdS
space without a brane (see, for example, \cite{bw1,bw2,bw3}). In
this particular case the branes in AdS-Schwarzschild are
conjectured to be dual to a radiation dominated FRW universe,
which is exactly the setup that we are interested in.  As we will
show, the maximal temperature $T_{max}$ has a geometric 5D
interpretation: it corresponds to the brane ``just" reaching  the
BH horizon.

The conjectured duality between branes propagating in
AdS-Schwarzschild space and a radiation dominated FRW universe
offers a novel perspective for studying the saturation of the
entropy bounds at $T_{MAX}$. The issue becomes whether the brane
can continue to fall into the BH and continues to be dual to a CFT
in a cosmological background at temperatures above $T_{MAX}$.

A possible way of viewing the propagation of branes in
AdS-Schwarzschild is the following: a thermal system with a known
form of entropy is thrown into a BH, a process analogous to the
Geroch process. Here a 4D universe is thrown whole into a 5D BH,
and so issues concerning the generalized second law (GSL) and its
relation to entropy bounds can be addressed. As in the standard
case, it is then possible to compare the total entropy of the
system before and after and to discuss cases in which a decrease
in the total entropy is suspected.  We do indeed find that the GSL
is violated as the brane falls into the BH.

In section~\ref{4D} we explain the saturation of entropy bounds at
$T_{MAX}$, and discuss possible physical mechanisms that may lead
to this saturation. In section~\ref{5D} we discuss the issue from
a 5D perspective, and discuss possible physical mechanisms that
may alter the propagation of branes with respect to naive
expectations. In section \ref{discussion} we offer several
possible resolutions of the puzzle that we have posed in the
previous sections.

\section{Black hole creation and a maximal temperature in four dimensions}
\label{4D}

Consider a relativistic gas in thermal equilibrium at a
temperature $T$. We assume that the gas consists of \N independent
degrees of freedom in a box of macroscopic linear size $R$, we
further assume that $R$ is larger than any fundamental length
scale in the system, and in particular R is much larger than the
Planck length $R\gg l_P$. The volume of the box is $V=R^3$. Since
the gas is in thermal equilibrium its energy density is $\rho=\N
T^4$ and its entropy density is $s=\N T^3$ (here and in the
following we systematically neglect numerical factors). As
explained previously, we are interested in the limit of large \N.

Under what conditions is this relativistic gas unstable to the
creation of BH's? The simplest criterion which may be used to
determine whether an instability is present is a comparison of the
total energy in the box $E_{\rm Th}=\N T^4 R^3$ to the energy of a
BH of the same size $E_{\rm BH}=M_P^2 R$. The two energies are
equal when $T^4= 1/\N M_P^2/R^2$. So thermal radiation in a box
and a BH of the same size have the same energy if
\begin{equation}
\label{energy} (TR)^4=\frac{1}{\N} M_P^2  R^2.
\end{equation}
Another criterion that may help us to determine the presence of an
instability to BH's creation is to compare the thermal entropy
$S_{\rm Th}=\N T^3 R^3$ to the entropy of the BH $S_{\rm BH}=M_P^2
R^2$. They are equal when $T^3= 1/\N M_P^2/R$. So thermal
radiation in a box and a BH of the same size have the same entropy
if
\begin{equation}
\label{entropy}
(TR)^3=1/\N M_P^2 R^2.
\end{equation}
From eqs. (\ref{energy}) and (\ref{entropy}) it is possible to
conclude the well known fact that for fixed $R$ and \N, if the
temperature is low enough the average thermal energy is not
sufficient to form BH's. For low temperatures the thermal
fluctuations are weak and they do not alter the conclusion
qualitatively.

Here we are interested in the case $RT>1$ which means that the
size of the box is larger than the thermal wavelength $1/T$. The
case $RT<1$ has been considered previously in \cite{shortest}. In
this case the temperature is not relevant. Instead, the field
theory cutoff $\Lambda$ was shown to be the relevant scale. In
\cite{shortest} we found a relationship between $\Lambda$,
$M_{Pl}$ and the number of fields \N which is somewhat different
than what we find here between $T$, $M_{Pl}$ and \N.

Imagine raising the temperature of the radiation from some low
value for which condition (\ref{energy}) is not satisfied to
higher and higher values such that eventually condition
(\ref{energy}) is saturated. Note that since $TR>1$ eq.
(\ref{energy}) is saturated before eq. (\ref{entropy}). We assume
that the size of the box $R$ is fixed during this process (the
number of species \N is also fixed), and estimate the backreaction
of the radiation energy density on the geometry of the box to
determine whether the assumption that the geometry of box is fixed
is consistent. To obtain a simple estimate we assume that the box
is spherical, homogeneous and isotropic. Then its expansion or
contraction rate is given by the Hubble parameter $H= \dot R/R$,
which is determined by the $00$ Einstein equation $H^2 M_{P}^2=\N
T^{4}$. However, if eq.(\ref{energy}) is satisfied then
$\frac{1}{R^2} M_{P}^2=\N T^{4}$, and therefore $HR\sim 1$. The
conclusion is that if eq. (\ref{energy}) is saturated then the
gravitational time scale is comparable to the light crossing time
of the box, and therefore it is inconsistent to assume that the
box has a fixed size which is independent of the energy density
inside it.

The conclusion from the previous discussion is that we need to
study the issue of stability or instability to the creation of
BH's in a box filled with thermal radiation in a time-dependent
setting, namely, in a cosmological setting, where
\begin{eqnarray}\label{ein1}
H^2 M_{P}^2=\N T^{4}\, .
\end{eqnarray}

Entropy bounds such as the Hubble entropy bound and others are
saturated if $S_{Th}=\N T^3 H^{-3}=S_{BH}=M_{Pl}^2 H^{-2}$. From
eq.~(\ref{ein1}) we understand that this happens for
\begin{eqnarray}
H=T\, ,\quad T=T_{MAX}=\frac{M_{P}}{\sqrt{\N}}\, .
\end{eqnarray}
Let us examine in more detail the physics of a radiation dominated
(RD) universe at temperatures near $T_{MAX}$. If $H=T$, the
cosmological horizon size $H^{-1}$ becomes comparable to the
wavelength of a typical particle of the relativistic gas,
$\lambda\sim T^{-1}$. If we go beyond this temperature, the
classical description of the particles that compose the gas in
terms of a homogeneous and isotropic fluid is no longer
appropriate, and thus neither is eq.~(\ref{ein1}). Alternatively,
one can think of $T_{MAX}$ as the temperature at which the Jeans
length of a typical thermal fluctuation becomes comparable to the
thermal wavelength, thus suggesting, again, that the approximation
of the gas by a homogeneous and isotropic fluid becomes
inappropriate. Yet another way to think about $T_{MAX}$ is as the
temperature at which the entropy within a thermal wavelength
becomes comparable to the entropy of a BH of the same size, thus
making BH entropically favored over single particle excitations.
Similarly, at $T=T_{MAX}$ the thermal energy inside a ``box" of
size $H^{-1}$, $E=\N T^4 H^{-3}$ is equal to the energy of a BH of
the same size, and also the free energies of both states become
comparable. All of the above supports the qualitative conclusion
that a state of \N degrees of freedom cannot be in thermal
equilibrium at temperatures above $T_{MAX}$. We will try to
examine this issue more quantitatively below.

Let us first close some possible loopholes in our analysis. One
possible loophole could have been if thermal fluctuations were too
large. This is not the case. The ratio of the energy in thermal
fluctuations,
\begin{eqnarray}
\label{fluct}
 \frac{\Delta E^2}{E^2}=\frac{1}{\N}\frac{1}{T^3 R^3}\, ,
\end{eqnarray}
is small compared to the average value of the energy in this
regime and is much smaller than unity for $RT>1$, and $\N \gg 1$.
Another possible loophole could have been, as in \cite{shortest},
a clash with the assumption that the semiclassical treatment is
valid. Since, in the case at hand, the energy is dominated by the
mean value of $\rho$, and not by the fluctuations, we do not have
problems with black hole evaporation: in fact it turns out that
for
\begin{eqnarray}
T\leq T_{C}=\sqrt{\frac{640\pi}{\N}}M_{P}
\end{eqnarray}
BH's can be treated classically and, as can be seen by inserting
the correct numerical factors into the definition of $T_{MAX}$,
$T_{MAX}<T_{C}$.

In any case, it is clear that at $T_{MAX}$ a significant change
must occur in the way we describe thermal equilibrium and the
assumption that we can treat gravity as semiclassical, only
providing matter with a geometric background. All these
considerations are well known in the case $\N=1$; but if $\N$ is a
large number the relevant scale can be much smaller than the
Planck scale.

We would like to discuss the issues in a more quantitative way. We
would like to estimate the time scale for the collapse of
perturbations which, if frequent and strong enough, will lead to
production of black holes. The perturbation equations which govern
their evolution are well known \cite{Mukhanov:1990me}; we present
here the equation governing the dynamics of the Bardeen potential
$\Phi$, in longitudinal gauge
\begin{eqnarray}
6 \ddot{\Phi} + 24{\mathcal H}\dot{\Phi} + 12(\dot{\mathcal
  H}+{\mathcal H}^2)\Phi - 2 \Delta \Phi=0\, ,
\end{eqnarray}
with ${\mathcal H}\equiv\dot{R}/R$, the dot denoting the
derivative with respect to conformal time $\eta$, and $\Delta$ the
spatial Laplacian operator. The solution of the perturbation
equations is quite standard. First, by means of the spatial
Fourier transform the Laplacian operator is expressed in terms of
the comoving wavenumber $k$ as $\Delta\rightarrow -k^2$. Then one
notices that, since  the background evolves as a power law in
conformal time, and, in particular, for a radiation dominated
contracting universe one has $R(\eta)\sim -\eta$, with
$-\infty<\eta<0$, the solution for the mode $\Phi_k$ can be
expressed in terms of the variable $x\equiv k \eta$ as:
\begin{eqnarray}
\Phi_k (\eta)=A_k F(x)\, ,
\end{eqnarray}
where $F(x)$ (whose explicit form is not needed here) scales as
$x^{-2}$ for $x\rightarrow-\infty$, diverges as $x^{-3}$ for
$x\rightarrow 0$, and is of order one for $x\sim -1$.

The factor $A_k$ can be determined through the perturbed Friedmann
equation which gives a relation between the Bardeen potential and
the density perturbation:
\begin{eqnarray}
\Phi_k (\eta)=\frac{3}{2 x^2}\frac{\delta \rho_k (x)}{\rho}\, .
\end{eqnarray}
We now observe that the thermal energy fluctuations are dominated
by the comoving wavenumber $k_T\equiv R T$ since the higher modes
are exponentially suppressed in the Boltzmann distribution, and
that we can estimate them via eq.~(\ref{fluct}). We further
observe that at the time $\eta_{MAX}$ when the critical
temperature $T_{MAX}$ is attained, one has $x=-1$ for the mode
that dominates the fluctuations. By combining all these elements
we may express the Bardeen potential at $\eta_{MAX}$ in terms of
an initial thermal fluctuation at some early time $\eta_{i}$:
\begin{eqnarray}
\Phi_{k_T} (\eta_{MAX})=\frac{3}{2}\frac{1}{\sqrt{\N}(RT)^{3/2}}\frac{1}
{x_i^2 F(x_i)}F(-1)\, .
\end{eqnarray}
The factor $x_i^2 F(x_i)$ is of order one, and so is $F(-1)$; this
leads us to the conclusion that the Bardeen potential is still
small at the critical time, due to the large factor
$\sqrt{\N}(RT)^{3/2}$ in the denominator.

To summarize, we have found that if the initial perturbations are
provided by thermal fluctuations, then their initial amplitude is
very small, and since they grow only as a power law, they do not
have enough time to become large before the critical temperature
is reached. We conclude that BH production from thermal
perturbations is not quick enough, so entropy bounds seem to be
violated.

At this point we cannot proceed further with semiclassical methods
and get a better idea on the state of a system when the
temperature is increased beyond $T_{MAX}$, or even whether this is
possible at all.

\section{Feeding a 4D  Braneworld to a 5D black hole  }
\label{5D}

We can gain some insight about the meaning of $T_{MAX}$, and
perhaps some further technical control by modelling a 4D RD universe
as a brane moving in an AdS$_{5}$-Schwarzschild spacetime.

For precision, we will take the following representation for the
bulk spacetime
\begin{eqnarray}
 \label{lnelem}
{\rm d}s^2=-H(R){\rm d}t^2+ \frac{1}{H(R)}{\rm d}R^2 + R^2{\rm
d}\Omega_3^2\, ,
\end{eqnarray}
where $H(R)=1+\frac{R^2}{L^2} - \frac{b^4 L^2}{R^2}$ vanishes at the black hole
horizon $R_{H}$ and $b=\left(\frac{8 G_N^{(5)}}{3 \pi}\frac{M}{L^2}\right)^{1/4}$,
$M$ being the black hole mass. $L$ is related to the cosmological constant of the
AdS and also to the brane tension $\lambda$, which is tuned in such a way as to
make a vanishing effective cosmological constant on the brane. Note that the line
element in eq.(\ref{lnelem}) describes only the part of spacetime outside the BH
horizon; this will become important and relevant shortly when we discuss the fate
of a brane that is about to fall into the BH.

For the AdS/CFT correspondence to be valid, $b$ must be large
$b\gg 1$ \cite{Witten:1998zw}, that is, the black hole must be
large and hot compared to the surrounding $AdS_5$. In this limit
the closed 4D universe can be treated as flat, and we can write
$R_H\simeq b L$, and $b\simeq\pi LT_0$, where $T_0$ is the Hawking
temperature of the black hole.

The motion of the brane through the bulk spacetime is viewed by a
brane observer as a cosmological evolution. According to the
prescription of the RS II model \cite{Randall:1999vf}, the 4D
brane is placed at the $Z_2$ symmetric point of the orbifold. On
the other hand, in the so called mirage cosmology
\cite{Kehagias:1999vr}, the brane is treated as a test object
following a geodesic motion. In both cases the evolution of the
brane in the AdS$_{5}$-Schwarzschild bulk mimics a FRW radiation
dominated cosmology. Thus, both prescriptions are useful for our
purposes. We will keep them in mind in the following discussion.

The brane can be described by its radial position as a function of
the proper time of the brane $R_{b}(\tau)$. The evolution of
$R_{b}(\tau)$ is determined by an effective Friedmann equation:
\begin{eqnarray}
\label{brane}
\left(\frac{\dot{R_{b}}}{R_{b}}\right)^2=\frac{b^4 L^2}{R_{b}^4} -
\frac{1}{R_{b}^2}\, ,
\end{eqnarray}
where the dot here stands for a derivative with respect to cosmic
time. Since, as we recall, $b\gg 1$ so that the curvature term is
always negligible, we ignore it in the following.
Eq.~(\ref{brane}) expresses the dynamics of the brane in terms of
5D quantities; we now focus on the case of a {\it contracting}
brane and translate those quantities into 4D ones in order to be
able to compare eq.~(\ref{brane}) with eq.~(\ref{ein1}).

The AdS/CFT correspondence tells us that the number of species in
the CFT is given by $\N = L^3/G_{N}^{(5)}$, while the 4D and the
5D Newton's constants are related by $L G_{N}^{(4)}=G_{N}^{(5)}$
(again, we consistently ignore numerical factors). This is enough
to make a comparison between eq.~(\ref{brane}) and
eq.~(\ref{ein1}) and to obtain the temperature measured on the
brane as $T=b/R_{b}$, which is also in accordance with the AdS/CFT
correspondence. In passing, we notice that one should not confuse
the temperature of the boundary CFT that is dual to the AdS bulk
theory with the Hawking temperature of the AdS BH as measured by a
bulk observer located at the coordinate $R$. The latter is given
by $T_0 /\sqrt{H(R)}$ and scales with $R$ in a similar way to the
CFT temperature only in the asymptotic limit $R\rightarrow
\infty$.

We now wish to see what happens in the 5D picture when the
limiting temperature is approached on the brane. By expressing
$M_P\equiv \sqrt{G_N^{(4)}}$ and $\N$ in terms of 5D quantities we
can see that $T_{MAX}\simeq 1/L$ and, since the corresponding
value for $R$ is $b/T_{MAX}$, we find that
\begin{eqnarray}
T\rightarrow T_{MAX}\quad\Longrightarrow\quad R_{b}\rightarrow
R_H\, .
\end{eqnarray}
$T_{MAX}$ is reached exactly when the brane reaches the BH horizon
and is about to enter into the black hole!

At this point the whole meaning of the AdS/CFT correspondence
becomes unclear since, as we have noted before, it is valid only
for the region outside the BH horizon. In the RS II picture the
brane represents the boundary of the bulk spacetime, which means
that beyond the brane at $R>R_{b}$ there is an identical copy of
the bulk spacetime at $R<R_{b}$. So, as the brane reaches and then
crosses the horizon, namely after $R_b$ (the position of the
brane) has become less than $R_H$ (the horizon), the 5D spacetime
is described by two identical copies of the interior region of an
AdS-Schwarzschild space, cut at $R=R_b$ and glued together, with
the brane placed at the $Z_2$ symmetry point. At this point the 5D
BH disappears, and it is not clear which brane CFT should be the
dual of the bulk theory.

One could imagine avoiding confusion about the interpretation of
the AdS/CFT correspondence when the brane reaches the BH horizon
 if the brane motion is interpreted according to the
``mirage" prescription. In this case, the brane is not the
boundary of spacetime. Rather, it is a probe brane moving through
the fixed bulk background. This approach is slightly more helpful
in our case. A reasonable interpretation of what transpires at
horizon crossing is that the 4D universe simply ends its existence
and disappears into the BH. The BH ``eats" the 4D universe, its
mass increases and so does its size, and entropy. Therefore the
final state from a 5D point of view is simply an
$AdS_5$-Schwarzschild space with a larger BH.

We are thus studying a process analogous to the Geroch process,
with the significant difference that, in the case at hand, an
entire universe is thrown into the BH. Therefore we can look at
the entropy balance during the process and see whether the GSL is
respected or not.

In order to have a vanishing effective cosmological constant on
the brane, one has $G_N^{(5)} \lambda \simeq L^{-1}$; this means
that at horizon crossing the total energy of the brane is
\begin{eqnarray}\label{enebra}
\left. E\right|_{R=R_H}\simeq\frac{b^3 L^2}{G_N^{(5)}}\, .
\end{eqnarray}
Comparing $E$ to $M\simeq \frac{b^4 L^2 }{G_N^{(5)}} $ we see that for $b\gg 1$ the
total energy of the brane is much smaller than the BH mass $E\ll M$.

The entropy of the 5D black hole is $S={\cal A}(R_H)/4 G_N^{(5)}$,
with the area of the horizon given by ${\cal A}(R_H)=2 \pi^2
R_H^3$. When the brane falls into the BH, the entropy of the BH is
increased by the following amount:
\begin{eqnarray}
\delta S\simeq\frac{1}{4 G_N^{(5)}}E\frac{\delta {\cal
A}(R_H(M))}{\delta M}\simeq \frac{E L^2 }{R_H}\simeq \frac{E L
}{b}\, .
\end{eqnarray}

For the GSL to hold, the total entropy of the system should
increase in the process
\begin{eqnarray}
\delta S>S_b\, .
\end{eqnarray}
Since $S_b = 2 \pi^2 R_H^3 {\cal N} T^3\simeq E L$ is the total
entropy on the brane when it is about to fall into the BH, we find
that for the total entropy to increase
\begin{eqnarray}
b<1\, .
\end{eqnarray}
However, for the AdS/CFT correspondence to hold, $b$ has to be
much larger than unity $b\gg 1$! If indeed $b\gg 1$, then
apparently the GSL is violated in this process. We have thus found
that a violation of the GSL in the 5D bulk corresponds to a
violation of the entropy bounds in the 4D brane. The situation is
completely analogous to the one discussed in connection with the
ordinary Geroch process where the GSL is apparently violated if
the falling object does not satisfy the Bekenstein bound. This
issue has a long history (see, for example, \cite{beb2},
\cite{Unruh:ic}-\cite{Marolf:2002ay}) and is controversial to some
extent. We do not attempt to take sides in the debate, but rather
to simply point out the similarities.

In any case, if the previous interpretations are correct, we must
conclude that the AdS/CFT correspondence seems to be incapable of
describing a RD universe in thermal equilibrium at temperatures
above $T_{MAX}$; in the mirage approach this happens because the
brane that hosts the CFT disappears, while in the RS II picture it
is the other way around: the 5D BH ceases its existence. In both
cases, this sudden breaking of the correspondence lends support to
the significance of $T_{MAX}$ as a temperature above which thermal
equilibrium physics is altered.

We may try to use the 5D picture to understand in a more
qualitative way what is the physical mechanism that renders
$T_{MAX}$ a limiting temperature. Black hole creation and the
subsequent ``breaking" of the brane seemed to be one of the
possibilities in the 4D picture. From the brane world point of
view this would correspond to the formation of ``blisters'' on the
brane. In fact, since the temperature of the brane scales as
$1/R_{b}$, if a piece of brane is closer to the BH with respect to
the rest of the brane, then the local temperature on that piece
will be higher, as will its energy density. A piece of the brane
that has higher energy density has a higher local magnitude of the
Hubble parameter. Therefore the speed at which it falls towards
the BH is increased, and we expect a ``blister" to form on the
brane. Thus a local oscillation of the brane position would be
seen by a brane observer as a local density perturbation which is
further amplified as the brane falls towards the BH. This
mechanism can be studied by looking at perturbation equations for
the position of the brane. Since these are coupled to the bulk
metric perturbations of the AdS$_{5}$, it seems that the full set
of perturbation equations must be studied.

However, as it turns out, in our case one can study the
perturbations directly from the 4D  brane point of view: it is
sufficient to write down the projected Einstein equations on the
brane as
\begin{eqnarray}\label{einbra}
G_{\mu\nu}=-E_{\mu\nu}\, ,
\end{eqnarray}
$E_{\mu\nu}$ being the projected bulk Weyl tensor on the brane
(see \cite{Shiromizu:1999wj}), and to perturb them.
Eq.~(\ref{einbra}) looks so simple because there is no matter on
the brane, just the tension which is fine-tuned in order to cancel
the bulk cosmological constant, so that both disappear from the
dynamics. The only effective source term is then the projection of
the bulk Weyl tensor, which we parameterize as a fluid with energy
density $\rho_{\epsilon}$ and pressure
$p_{\epsilon}=-\rho_{\epsilon}/3$ ($E_{\mu\nu}$ is traceless).
Thus the system of perturbation equations is closed and can be
solved without reference to the 5D picture. Notice that this
happens because of the simplicity of the model at hand: if we had
some matter on the brane this would no longer have been true.

In the end, the perturbation equations look exactly the same as in
the pure 4D scenario discussed in the previous section and the
same physical considerations about the growth of perturbations are
valid. So it seems that the standard picture is confirmed: as
$R\rightarrow R_H$, $H\rightarrow T$, and at horizon crossing the
typical modes in the thermal bath become unstable. However their
growth follows a power law only, and thus there is not enough time
for the instability to invalidate the whole picture.

Another possible 5D mechanism that could modify our discussion and
its conclusion about the saturation of the entropy bounds is the
interaction of the bulk Hawking radiation with the brane. Since,
as we have seen, the temperature of the Hawking radiation diverges
at the horizon, one might have expected that at some point the
Hawking radiation pressure becomes so high that it prevents the
brane from falling into the BH. Perhaps the Hawking radiation
pressure could cause the brane to bounce back and change its
contraction into expansion or cause it to float just above the
horizon. We think that this is unlikely. However, clearly the
issue deserves further study, especially in light of the fact that
for $b\gg1$ the temperature of the Hawking radiation is also very
large. For boxes falling into BH's the issue was debated
extensively in the context of the relationship between the GSL and
entropy bounds \cite{Unruh:ic}-\cite{Marolf:2002ay}.

We would like to make a few observations about the possible
influence of the Hawking radiation on the motion of the brane.

First, in AdS space the geometry provides a confining environment
for the radiation which is then in equilibrium with the BH. Notice
also that, unlike the pure Schwarzschild case, here the
equilibrium is stable. The pressure on the brane results from the
difference in the force exerted on the two sides of the brane. If
the system is in thermal equilibrium and the brane is moving
through the radiation fluid, then the pressure on it depends on
the interaction of the brane with the radiation. If it is
transparent, then the radiation does not exert any pressure on the
brane, and if it is opaque, then the radiation pressure can be
estimated by the pressure of a fluid at the Hawking temperature.

The Hawking temperature at the position of the brane is given by
$T_H=T_0/\sqrt{H(R_b)} \simeq \frac{b}{L \sqrt{H(R_b)}}$.
Substituting $H(R_b)=1+\frac{R_b^2}{L^2}\left(1-\frac{b^4
L^4}{R_b^4}\right)$, we see that as long as the distance of the
brane from the horizon $R_b-bL$ remains finite, then $T_H\sim
b/R_b \sim T_{brane}$. We then observe that the Hawking radiation
pressure is smaller by a factor of \N compared to the pressure on
the brane, which in turn determines the acceleration of the brane
towards the BH. We conclude that as long as the distance of the
brane from the horizon is not particularly small, the Hawking
radiation pressure is not likely to alter its motion
significantly.

When the brane does get close to the BH it seems that the Hawking
radiation pressure can affect the motion of the brane. However, it
is not clear whether the fluid description of the Hawking
radiation is valid in the vicinity of the horizon. The wavelength
of a typical particle in thermal bath at temperature T is
$\lambda\sim T^{-1}$, and the typical wavelength of the Hawking
radiation in our AdS-Schwarzschild spacetime is
\begin{eqnarray}
\lambda_{H}(R)\sim \frac{\pi L}{b}\sqrt{\frac{R^2}{L^2}-
\frac{b^4 L^2}{R^2}}\, ,
\end{eqnarray}
where we have taken into account the behavior of the local Hawking
temperature as discussed above.

On the other hand the physical distance of a spacetime point
with radial coordinate R from the horizon is
\begin{eqnarray}
d(R)=\int_{b L}^{R}\sqrt{g_{RR}(x)}{\rm
  d}x=\frac{L}{2}\log{\left[\left(\frac{R}{b L}\right)^2 +
\sqrt{\left(\frac{R}{b L}\right)^4 -1}\right]}\, .
\end{eqnarray}
Notice that although $g_{RR}$ diverges near the horizon, $d(R)$ is
always finite at finite $R$.

Now observe that as one gets close to the horizon (i.e. for
$R-bL\ll bL$), the following relation holds
\begin{eqnarray}
\lambda_{H}(R)\rightarrow 2 \pi d(R)\, ,
\end{eqnarray}
meaning that the typical wavelength $\lambda_{H}$ becomes larger
than (or in any case, of the same order of magnitude as)  the
physical distance from the horizon, thus implying that the
description of the Hawking radiation as a fluid becomes
inappropriate at this point. One could then argue that the Hawking
radiation forms mostly at distances $d \sim bL$ from the black
hole and larger, and that for smaller distances there is no
significant radiation pressure. This means that Hawking radiation
pressure cannot stop the brane from falling into the BH as it
approaches the horizon.

These issues were discussed in the context of falling boxes most
recently by Marolf and Sorkin \cite{Marolf:2002ay}, and previously
by others. We conclude that the answer depends on the detailed
dynamics of the system.

\section{Discussion and possible resolution}
\label{discussion}

We have seen that a special value of the temperature
$T_{MAX}=M_{P}/\sqrt{\N}$ emerges in various contexts. We have
seen that such a value arises in four dimensional models as the
temperature at which entropy bounds are saturated, and in five
dimensional models as the effective induced temperature on a brane
propagating in AdS-Schwarzschild spacetime as it reaches the
horizon of the bulk BH and is about to disappear into it. We have
also shown that in the five dimensional picture the GSL is
violated as the brane falls into the BH.

We have presented some examples for the appearance of this special
value of the temperature, and have provided arguments supporting
its existence or that a change in the description of equilibrium
physics at this temperature is required. We have not provided
conclusive evidence as to whether a specific physical mechanism is
responsible for enforcing such a maximal temperature, or whether
one exists at all. We have not been able to identify a single
mechanism that is efficient and quick enough to  prevent the
universe from exceeding  the limiting temperature nor to identify
the required changes in the description of physics at this
temperature.

We list a few possibilities which we leave as unsolved puzzles and
 interesting problems for future research:

\begin{enumerate}

\item

Entropy bounds give the correct limiting temperature in their
currently known form. Some enforcing mechanism exists which is
still unknown.

\item

Entropy bounds need to be modified such that the limiting
temperature disappears, and they are consistent at all
temperatures.

\item

Brane world AdS/CFT correspondence is valid for the model that we
are considering: that the brane falls into the BH and entropy
bounds are violated.

\item

Brane world AdS/CFT correspondence  in this particular context is
not valid when the brane approaches the horizon and falls into the
BH. When modified appropriately, for example by correctly taking
into account the influence of the Hawking radiation pressure or
the growth of perturbations or the effects of additional induced
matter on the brane, entropy bounds remain valid in their
currently known forms.

\item

Both the AdS/CFT correspondence in this specific context and the
currently known entropy bounds are not valid for temperatures of
about $T_{MAX}$.

\item

The number of light fields \N is fundamentally limited, a fact
which is well represented by entropy bounds, and therefore
considering the large \N limit $\N \to \infty$, as is done in the
AdS/CFT correspondence, is incorrect.

\end{enumerate}

At this point in time we do not have a clear preference or a clear
indication from our calculations as to which of these
possibilities is correct. We hope that future research will help
to resolve the issues that we have discussed.

\section{acknowledgments}
This research was supported in part by the Israel Science
Foundation under grant no. 174/00-2 and by the NSF under grant no.
PHY-99-07949. S.~F. was partially supported by the Kreitman
foundation.  R.~B. thanks the KITP, UC at Santa Barbara, where
this work was completed. We thank the participants of the string
cosmology program at KITP for comments and D. Marolf in particular
for discussions and helpful suggestions.


\begin{thebibliography}{10}

\bibitem{beb1}
J.~D.~Bekenstein,
%``Is The Cosmological Singularity Thermodynamically Possible?,''
Int.\ J.\ Theor.\ Phys.\  {\bf 28}, 967 (1989).
%%CITATION = IJTPB,28,967;%%

\bibitem{beb2}
J.~D.~Bekenstein,
 %``A Universal Upper Bound On The Entropy To Energy Ratio For Bounded
%Systems,''
Phys.\ Rev.\ D {\bf 23}, 287 (1981);
%%CITATION = PHRVA,D23,287;%%
J.~D.~Bekenstein,
%``Entropy bounds and black hole remnants,''
Phys.\ Rev.\ D {\bf 49}, 1912 (1994) [arXiv:gr-qc/9307035].
%%CITATION = GR-QC 9307035;%%

\bibitem{veb}
G.~Veneziano,
%``Entropy bounds and string cosmology,''
Phys.\ Lett.\ {\bf B454}, 22 (1999)
[arXiv:hep-th/9907012].
%%CITATION = HEP-TH 9907012;%%

\bibitem{ch}
B.~J.~Carr and S.~W.~Hawking,
%``Black Holes In The Early Universe,''
Mon.\ Not.\ Roy.\ Astron.\ Soc.\  {\bf 168}, 399 (1974);
%%CITATION = MNRAA,168,399;%%
B.~J.~Carr,
%``The Primordial Black Hole Mass Spectrum,''
Astrophys.\ J.\  {\bf 201}, 1 (1975);
%%CITATION = ASJOA,201,1;%%

\bibitem{holo}
G.~'t Hooft,
%``Dimensional Reduction In Quantum Gravity,''
arXiv:gr-qc/9310026;
%%CITATION = GR-QC 9310026;%%
L.~Susskind,
%``The World as a hologram,''
J.\ Math.\ Phys.\  {\bf 36}, 6377 (1995)
[arXiv:hep-th/9409089].
%%CITATION = HEP-TH 9409089;%%

\bibitem{fs}
W.~Fischler and L.~Susskind,
%``Holography and cosmology,''
arXiv:hep-th/9806039.
%%CITATION = HEP-TH 9806039;%%

\bibitem{EL}
R.~Easther and D.~A.~Lowe,
%``Holography, cosmology and the second law of thermodynamics,''
Phys.\ Rev.\ Lett.\  {\bf 82}, 4967 (1999)
[arXiv:hep-th/9902088].
%%CITATION = HEP-TH 9902088;%%

\bibitem{KL}
N.~Kaloper and A.~D.~Linde,
%``Cosmology vs. holography,''
Phys.\ Rev.\ D {\bf 60}, 103509 (1999)
[arXiv:hep-th/9904120].
%%CITATION = HEP-TH 9904120;%%

\bibitem{entbound1}
D.~Bak and S.~J.~Rey,
%``Cosmic holography,''
Class.\ Quant.\ Grav.\  {\bf 17}, L83 (2000)
[arXiv:hep-th/9902173];
%%CITATION = HEP-TH 9902173;%%

\bibitem{entbound2}
R.~Bousso,
%``A Covariant Entropy Conjecture,''
JHEP {\bf 9907}, 004 (1999)
[arXiv:hep-th/9905177];
%%CITATION = HEP-TH 9905177;%%

\bibitem{entbound3}
R.~Bousso,
%``Holography in general space-times,''
JHEP {\bf 9906}, 028 (1999)
[arXiv:hep-th/9906022];
%%CITATION = HEP-TH 9906022;%%

\bibitem{entbound4}
R.~Bousso,
%``The holographic principle for general backgrounds,''
Class.\ Quant.\ Grav.\  {\bf 17}, 997 (2000)
[arXiv:hep-th/9911002];
%%CITATION = HEP-TH 9911002;%%

\bibitem{entbound5}
R.~Brustein and G.~Veneziano,
%``A Causal Entropy Bound,''
Phys.\ Rev.\ Lett.\  {\bf 84}, 5695 (2000)
[arXiv:hep-th/9912055].
%%CITATION = HEP-TH 9912055;%%

\bibitem{entbound6}
E.~Verlinde,
%``On the holographic principle in a radiation dominated universe,''
arXiv:hep-th/0008140;
%%CITATION = HEP-TH 0008140;%%

\bibitem{entbound7}
I.~Savonije and E.~Verlinde,
%``CFT and entropy on the brane,''
Phys.\ Lett.\ B {\bf 507}, 305 (2001)
[arXiv:hep-th/0102042].
%%CITATION = HEP-TH 0102042;%%

\bibitem{gsl}
R.~Brustein,
%``The generalized second law of thermodynamics in cosmology,''
Phys.\ Rev.\ Lett.\  {\bf 84}, 2072 (2000)
[arXiv:gr-qc/9904061].
%%CITATION = GR-QC 9904061;%%m{gsl}

\bibitem{gslstring}
R.~Brustein, S.~Foffa and R.~Sturani,
%``Generalized second law in string cosmology,''
Phys.\ Lett.\ B {\bf 471}, 352 (2000)
[arXiv:hep-th/9907032].
%%CITATION = HEP-TH 9907032;%%

%\cite{Brustein:2001di}
\bibitem{Brustein:2001di}
R.~Brustein, S.~Foffa and G.~Veneziano,
%``CFT, holography, and causal entropy bound,''
Phys.\ Lett.\ B {\bf 507}, 270 (2001) [arXiv:hep-th/0101083].
%%CITATION = HEP-TH 0101083;%%


\bibitem{Maldacena:1997re}
J.~M.~Maldacena,
%``The large N limit of superconformal field theories and
%supergravity,''
Adv.\ Theor.\ Math.\ Phys.\  {\bf 2}, 231 (1998)
[Int.\ J.\ Theor.\ Phys.\  {\bf 38}, 1113 (1999)]
[arXiv:hep-th/9711200].
%%CITATION = HEP-TH 9711200;%%

\bibitem{adsrev}
O.~Aharony, S.~S.~Gubser, J.~M.~Maldacena, H.~Ooguri and Y.~Oz,
%``Large N field theories, string theory and gravity,''
Phys.\ Rept.\  {\bf 323}, 183 (2000) [arXiv:hep-th/9905111].
%%CITATION = HEP-TH 9905111;%%

\bibitem{Kraus:1999it}
P.~Kraus,
%``Dynamics of anti-de Sitter domain walls,''
JHEP {\bf 9912}, 011 (1999) [arXiv:hep-th/9910149].
%%CITATION = HEP-TH 9910149;%%

\bibitem{Kehagias:1999vr}
A.~Kehagias and E.~Kiritsis,
%``Mirage cosmology,''
JHEP {\bf 9911}, 022 (1999) [arXiv:hep-th/9910174].
%%CITATION = HEP-TH 9910174;%%

\bibitem{Ida:1999ui}
D.~Ida,
%``Brane-world cosmology,''
JHEP {\bf 0009}, 014 (2000) [arXiv:gr-qc/9912002].
%%CITATION = GR-QC 9912002;%%

%\cite{Wang:2001bf}
\bibitem{Wang1}
B.~Wang, E.~Abdalla and R.~K.~Su,
 %``Relating Friedmann equation to Cardy formula in universes with  cosmological
%constant,''
Phys.\ Lett.\ B {\bf 503}, 394 (2001) [arXiv:hep-th/0101073].
%%CITATION = HEP-TH 0101073;%%



%\cite{Wang:2001bv}
\bibitem{Wang2}
B.~Wang, E.~Abdalla and R.~K.~Su,
%``Friedmann equation and Cardy formula correspondence in brane universes,''
Mod.\ Phys.\ Lett.\ A {\bf 17}, 23 (2002) [arXiv:hep-th/0106086].
%%CITATION = HEP-TH 0106086;%%



\bibitem{Witten:1998zw}
E.~Witten,
%``Anti-de Sitter space, thermal phase transition, and confinement in
% gauge theories,''
Adv.\ Theor.\ Math.\ Phys.\  {\bf 2}, 505 (1998)
[arXiv:hep-th/9803131].
%%CITATION = HEP-TH 9803131;%%

\bibitem{Gubser:1999vj}
S.~S.~Gubser,
%``AdS/CFT and gravity,''
Phys.\ Rev.\ D {\bf 63}, 084017 (2001)
[arXiv:hep-th/9912001].
%%CITATION = HEP-TH 9912001;%%


%\cite{Anchordoqui:2001qc}
\bibitem{bw1}
L.~Anchordoqui, J.~D.~Edelstein, C.~Nunez, S.~E.~Perez Bergliaffa,
M.~Schvellinger, M.~Trobo and F.~Zyserman,
%``Brane worlds, string cosmology, and AdS/CFT,''
Phys.\ Rev.\ D {\bf 64}, 084027 (2001) [arXiv:hep-th/0106127].
%%CITATION = HEP-TH 0106127;%%

%\cite{Nojiri:2002hz}
\bibitem{bw2}
S.~Nojiri, S.~D.~Odintsov and S.~Ogushi,
 %``Friedmann-Robertson-Walker brane cosmological equations from the
%five-dimensional bulk (A)dS black hole,''
Int.\ J.\ Mod.\ Phys.\ A {\bf 17}, 4809 (2002)
[arXiv:hep-th/0205187].
%%CITATION = HEP-TH 0205187;%%

%\cite{Maartens:2003tw}
\bibitem{bw3}
R.~Maartens,
%``Brane-world gravity,''
arXiv:gr-qc/0312059.
%%CITATION = GR-QC 0312059;%%


\bibitem{shortest}
R.~Brustein, D.~Eichler, S.~Foffa and D.~H.~Oaknin,
%``The shortest scale of quantum field theory,''
Phys.\ Rev.\ D {\bf 65}, 105013 (2002)
[arXiv:hep-th/0009063].
%%CITATION = HEP-TH 0009063;%%

\bibitem{Mukhanov:1990me}
V.~F.~Mukhanov, H.~A.~Feldman and R.~H.~Brandenberger,
 %``Theory Of Cosmological Perturbations. Part 1. Classical Perturbations.
Part
%2. Quantum Theory Of Perturbations. Part 3. Extensions,''
Phys.\ Rept.\  {\bf 215}, 203 (1992).
%%CITATION = PRPLC,215,203;%%

\bibitem{Randall:1999vf}
L.~Randall and R.~Sundrum,
%``An alternative to compactification,''
Phys.\ Rev.\ Lett.\  {\bf 83}, 4690 (1999)
[arXiv:hep-th/9906064].
%%CITATION = HEP-TH 9906064;%%

\bibitem{Shiromizu:1999wj}
T.~Shiromizu, K.~i.~Maeda and M.~Sasaki,
%``The Einstein equations on the 3-brane world,''
Phys.\ Rev.\ D {\bf 62}, 024012 (2000)
[arXiv:gr-qc/9910076].
%%CITATION = GR-QC 9910076;%%






%\cite{Unruh:ic}
\bibitem{Unruh:ic}
W.~G.~Unruh and R.~M.~Wald,
%``Acceleration Radiation And Generalized Second Law Of Thermodynamics,''
Phys.\ Rev.\ D {\bf 25}, 942 (1982).
%%CITATION = PHRVA,D25,942;%%



%\cite{Li:sy}
\bibitem{Li:sy}
L.~X.~Li and L.~Liu,
 %``Properties Of Radiation Near The Black Hole Horizon And The Second Law Of
%Thermodynamics,''
Phys.\ Rev.\ D {\bf 46}, 3296 (1992).
%%CITATION = PHRVA,D46,3296;%%


%\cite{Bekenstein:1993dz}
\bibitem{Bekenstein:1993dz}
J.~D.~Bekenstein,
%``Entropy bounds and black hole remnants,''
Phys.\ Rev.\ D {\bf 49}, 1912 (1994) [arXiv:gr-qc/9307035].
%%CITATION = GR-QC 9307035;%%

%\cite{Anderson:pi}
\bibitem{Anderson:pi}
W.~G.~Anderson,
%``The Boulware State And The Generalized Second Law Of Thermodynamics,''
Phys.\ Rev.\ D {\bf 50}, 4786 (1994) [arXiv:gr-qc/9402030].
%%CITATION = GR-QC 9402030;%%

%\cite{Bekenstein:1999bh}
\bibitem{Bekenstein:1999bh}
J.~D.~Bekenstein,
 %``Non-Archimedean character of quantum buoyancy and the generalized  second
%law of thermodynamics,''
Phys.\ Rev.\ D {\bf 60}, 124010 (1999) [arXiv:gr-qc/9906058].
%%CITATION = GR-QC 9906058;%%

%\cite{Marolf:2002ay}
\bibitem{Marolf:2002ay}
D.~Marolf and R.~Sorkin,
%``Perfect mirrors and the self-accelerating box paradox,''
Phys.\ Rev.\ D {\bf 66}, 104004 (2002) [arXiv:hep-th/0201255].
%%CITATION = HEP-TH 0201255;%%



\end{thebibliography}
\end{document}